# Signature of long-ranged spin triplets across a two-dimensional superconductor/helimagnet van der Waals interface

A. Spuri[1], D. Nikolić[1,2], S. Chakraborty[1], M. Klang[3], H. Alpern[3], O. Millo[3], H. Steinberg[3], W. Belzig[1], E. Scheer[1], A. Di Bernardo[1,4*]

1. *Department of Physics, University of Konstanz, 78457 Konstanz, Germany.*
2. *Institut für Physik, Universität Greifswald, Felix-Hausdorff-Strasse 6, 17849 Greifswald, Germany*
3. *Racah Institute of Physics, The Hebrew University of Jerusalem, 91904 Jerusalem, Israel*
4. *Dipartimento di Fisica "E. R. Caianiello", Università degli Studi di Salerno, I-84084, Fisciano (SA), Italy*

*Correspondence to: angelo.dibernardo@uni-konstanz.de

**Abstract**
The combination of a superconductor with a magnetically inhomogeneous material has been established as an efficient mechanism for the generation of long-ranged spin-polarized (spin-triplet) Cooper pairs. Evidence for this mechanism, however, has been demonstrated based on studies done on thin film multilayers, where the strong bonds existing at the interface between the superconductor and the magnetic material should in principle enhance proximity effects and strengthen any electronic correlations. Here, we fabricate devices based on van der Waals (vdW) stacks of flakes of the $NbS_2$ combined with flakes of $Cr_{1/3}NbS_2$, which has a built-in magnetic inhomogeneity due to its helimagnetic spin texture at low temperatures. We find that the critical temperature of these vdW heterostructures is strongly dependent on the magnetic state of $Cr_{1/3}NbS_2$, whose degree of magnetic inhomogeneity can be controlled via an applied magnetic field. Our results demonstrate evidence for the generation of long-ranged spin-triplet pairs across the $Cr_{1/3}NbS_2/NbS_2$ vdW interface.

## Introduction

The interplay between ferromagnetism and superconductivity has been widely investigated in superconductor/ferromagnet (S/F) thin film multilayers with a conventional S, where the Cooper pairs of electrons are in an antiparallel-aligned (spin-singlet) state. Although early studies on such S/F hybrids showed that magnetic exchange field ($h_{ex}$) of a F quickly suppresses the spin-singlet superconductivity from the S/F interface [1-3], over the past 20 years several groups have demonstrated, confirming a theoretical prediction [4], that long-ranged parallel-aligned (spin-triplet) Cooper pairs of electrons can be generated in S/F devices with non-collinear Fs [5-9] or with an intrinsically-inhomogeneous magnetic material like the helimagnet Ho [10-12]. Evidence for long-ranged spin triplets in these S/F hybrids has been found via measurements of the dependence of their superconducting critical temperature ($T_c$) on the magnetic state (i.e., collinear/non-collinear) [7-9] or via measurements of a long-ranged supercurrents (compared to the F's coherence length, $\xi_F$, for spin-singlet pairs) in S/F/S Josephson junctions (JJs) [4-6,10,13]. Spectroscopic evidence for spin-triplet states has also been reported via measurements of sub-gap states in the superconducting density of states [12,14-15] or via measurements of an inverse Meissner effect [11]. All these studies have established the research field of superconducting spintronics [16] that aims at exploiting long-ranged spin triplets, which carry a net spin, to do spintronics in the superconducting state with low-energy dissipation.

Van der Waals (vdW) materials can be exfoliated down to the two-dimensional limit and exhibit physical properties that often do not have a counterpart in three-dimensional systems. Over the past years, a variety of vdW Ss and Fs has been explored both as single flakes and stacked to form vdW heterostructures [17-21]. S/F vdW heterostructures can be used to realize superconducting spintronic devices with novel functionalities compared to those based on thin film multilayers. Before this occurs, however, it is essential to determine at which S/F vdW interfaces long-ranged spin triplets can be generated

Several groups have recently studied the proximity effect in S/F vdW hybrids, and they have observed signatures of phenomena like a 0-to-π transition in S/F/S vdW JJs [22], which had only been known for JJs made from thin film multilayers [2-3]. Spin-triplet generation at the interface between electrodes of NbN (S) and the vdW material $Fe_{0.29}TaS_2$ (F) has also been reported [23], although this S/F interface may not be truly vdW in nature. A long-ranged coupling has also been measured in lateral $NbSe_2/Fe_3GeTe_2/NbSe_2$ (S/F/S) vdW JJs [24], but the spin-triplet nature of the supercurrent and the mechanism behind its generation remains unclear because $Fe_3GeTe_2$ does not have an intrinsic magnetic inhomogeneity.

In this Letter, we explore the superconducting proximity effect across the vdW interface forming between a flake of a vdW S ($NbS_2$) and a magnetic flake of $Cr_{1/3}NbS_2$. $Cr_{1/3}NbS_2$ is helimagnetic at low temperatures [26-27] and therefore has a built-in magnetic inhomogeneity ideal for spin-triplet generation. Studying the evolution of the $T_c$ of $Cr_{1/3}NbS_2/NbS_2$ (F/S) stacks as a function of the magnetic state of $Cr_{1/3}NbS_2$, we find that the $T_c$ of these vdW devices strongly depends on the $Cr_{1/3}NbS_2$ magnetization, in a way that cannot be reconciled with a conventional short-ranged S/F proximity effect or with stray fields. Supported also by a theoretical model, we show that our results are consistent with the generation of long-ranged spin-triplet pairs across the $Cr_{1/3}NbS_2/NbS_2$ vdW interface.

**Results and discussion**

We fabricate vdW heterostructures of $Cr_{1/3}NbS_2/NbS_2$ ($Cr_{1/3}NbS_2$ on top) on pre-patterned Au(33 nm)/Ti(7 nm) electrodes using the dry-transfer technique [25], as shown in Fig. 1(a). We use $Cr_{1/3}NbS_2$ flakes with thickness ranging between 200 nm and 500 nm, which are obtained via subsequent mechanical cleaving and exfoliation of $Cr_{1/3}NbS_2$ single crystals synthesized as in ref. [26].

We choose $Cr_{1/3}NbS_2$ as our magnetic flake because it is the closest equivalent to the helimagnet Ho, which previous studies on S/F thin film multilayers have shown to be efficient for spin-triplet generation [10-12]. Like in Ho, the helimagnetic spin texture of $Cr_{1/3}NbS_2$ can be unzipped by an in-plane magnetic field $H$ which, above a certain value depending on thickness [26-27], makes $Cr_{1/3}NbS_2$ fully ferromagnetic. This property provides a tool to control whether any observed effects is due to long-ranged spin triplets because such triplets should be only generated in the magnetically-inhomogeneous (helimagnetic) state of $Cr_{1/3}NbS_2$ and be suppressed when $Cr_{1/3}NbS_2$ is driven into its magnetically-homogeneous (ferromagnetic) state.

$Cr_{1/3}NbS_2$ is an ionic compound, which makes it difficult to obtain flakes with thickness smaller than 50 nm [26-27]. Nonetheless, this is not a limitation for our experiment, since we deliberately choose $Cr_{1/3}NbS_2$ flakes with thickness larger than 200 nm. Thick flakes of $Cr_{1/3}NbS_2$ are necessary because, if long-ranged spin-triplet pairs are generated at the $Cr_{1/3}NbS_2/NbS_2$ vdW interface, these pairs must have enough room to propagate into $Cr_{1/3}NbS_2$ for them to affect the $T_c$ of the proximitized $NbS_2$. Theory and experiments on S/F thin film multilayers have in fact shown that, to maximize the effect of long-ranged spin triplets on $T_c$, a F with thickness $d_F$ larger than its $\xi_F$ is needed [9, 28]. To increase the effect on $T_c$, the S should also have thickness $d_S$ comparable to or smaller than its superconducting coherence length ($\xi_S$) [8-9].

The reason why long-ranged spin triplets can reduce the $T_c$ of a S/F heterostructure with $d_S \sim \xi_S$ and $d_F > \xi_F$ can be understood by thinking of a S as a reservoir of Cooper pairs: once long-ranged spin-triplet pairs are generated, they can propagate deeply into the F (much deeper than $\xi_F$), which drains pairs out of the S and reduces $T_c$. On the other hand, if only short-ranged spin-triplet pairs are generated, the proximity effect remains confined at the S/F interface within $\xi_F$ (typically a few nanometers [29]) and $T_c$ is higher.

To determine whether long-ranged spin triplets are generated in our $Cr_{1/3}NbS_2/NbS_2$ devices, we study the evolution of their resistance versus temperature, $R(T)$, measured across $T_c$, as a function of an in-plane $H$ (i.e., perpendicular to the $Cr_{1/3}NbS_2$ c-axis). As $H$ is increased, the helimagnetic texture of $Cr_{1/3}NbS_2$ progressively unzips until $H$ reaches a saturation field ($H_{sat}$), at which $Cr_{1/3}NbS_2$ becomes ferromagnetic [26-27].

For the $Cr_{1/3}NbS_2/NbS_2$ device in Fig. 1(a), the $R(T)$ curve measured in $H = 0$ shows a peak-like feature, $R_{peak}$, at the onset of the superconducting transition. $R_{peak}$ persists in the helimagnetic state of $Cr_{1/3}NbS_2$ as $H$ is increased and it vanishes as $H$ approaches $\mu_0 H_{sat} \sim 0.5$ Tesla ($\mu_0$ being the vacuum permeability). At $H > H_{sat}$, $R_{peak}$ coincides with the normal-state resistance of the device ($R_N$) at $T \sim 6.0$ K (Figs. 1(b) and 1(c)). These observations and the fact that $R_{peak}$ is only measured in this device with a specific arrangement of contacts rule out magnetic impurities as explanation for $R_{peak}$. The data reported in Fig. 1(c) also show that, as the $H$ polarity is reversed and $H$ is decreased below $H_{sat}$, $R_{peak}$ reappears in the helimagnetic state of $Cr_{1/3}NbS_2$ before vanishing again at $H \sim -H_{sat}$. These results suggest that $R_{peak}$ must be correlated to the magnetically-inhomogeneous state of $Cr_{1/3}NbS_2$.

To understand the $R_{peak}$ origin, we note that, for the device in Fig. 1(a), we inject the current with two electrodes ($I^-$ and $I^+$) that are contacting only the bare $NbS_2$, and we collect the voltage signal with two other electrodes ($V^-$ and $V^+$) placed under the $Cr_{1/3}NbS_2/NbS_2$ stack. These two regions (i.e., bare $NbS_2$ and $Cr_{1/3}NbS_2/NbS_2$ stack) may have different $T_c$, which should be lower for the S/F stack due to the proximity effect [29]. As done in ref. [30], where a similar $R_{peak}$ is observed, in the Supplementary Material we show that $R_{peak}$ can be reproduced considering a network of resistors equivalent to our sample, where each resistor corresponds to a sample region with different $T_c$. Our analysis also confirms that the regions in the F/S stack have lower $T_c$ than those in the bare S.

For $R_{peak}$ to vanish at $|H| > H_{sat}$, the $T_c$ of the F/S stack ($T_{c,bi}$), which is lower than the $T_c$ of the bare S ($T_{c,bare}$) for $H < H_{sat}$, must become closer to $T_{c,bare}$ when $H > H_{sat}$ (Fig. 1(d)). This behavior of $T_{c,bi}$ is opposite to that expected for a F/S proximity with a homogeneous F or for stray fields, since both effects should get stronger and decrease $T_{c,bi}$ as $Cr_{1/3}NbS_2$ is driven ferromagnetic (see discussion below).

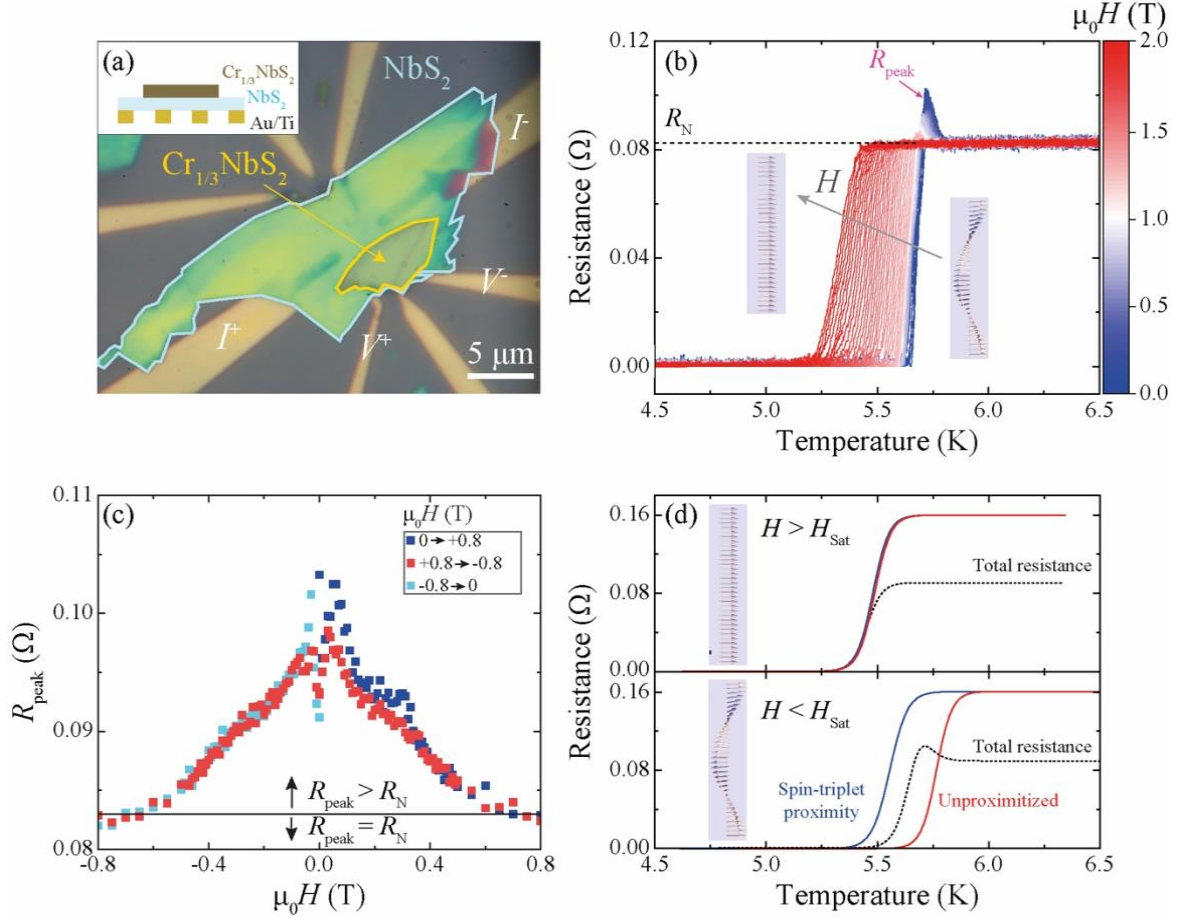

FIG. 1. (a) Optical microscope image of a $Cr_{1/3}NbS_2$ (200 nm)/$NbS_2$ (20 nm) vdW heterostructure on Au (23 nm)/Ti (7 nm) electrodes, with inset showing the materials stack, and corresponding $R(T)$ as a function of an in-plane $H$ in (b). The $R_{peak}$ feature vanishes as $H$ is increased above $H_{sat}$. (c) $R_{peak}$ versus $H$ for $H$ ranges specified in the legend. (d) Model showing the origin of $R_{peak}$ due to the presence of regions with different $T_c$ for $H < H_{sat}$ (bottom panel) assuming equal $T_c$ for $H > H_{sat}$ (top panel). Red and blue curves refer to bare $NbS_2$ and $Cr_{1/3}NbS_2/NbS_2$, respectively.

The positive shift of $T_{c,bi}$ towards $T_{c,bare}$ for $H > H_{sat}$ (Fig. 1(d)), which leads to the $R_{peak}$ disappearance can be due to an unconventional proximity effect at the $Cr_{1/3}NbS_2/NbS_2$ (F/S) vdW interface involving long-ranged spin triplets. This is because, when $Cr_{1/3}NbS_2$ is in its helimagnetic state, long-ranged spin triplets can be generated and can propagate deeply into $Cr_{1/3}NbS_2$ reducing $T_{c,bi}$ compared to $T_{c,bare}$ (Fig. 1(d); bottom). As $Cr_{1/3}NbS_2$ is driven ferromagnetic for $H > H_{sat}$, only short-ranged Cooper pairs can be generated, which remain confined close to the $Cr_{1/3}NbS_2/NbS_2$ interface making $T_{c,bi}$ approach $T_{c,bare}$ (Fig. 1(d); top).

The F/S vdW device in Fig. 1, however, makes it difficult to argue generation of long-ranged spin triplets across the $Cr_{1/3}NbS_2/NbS_2$ vdW interface due the presence of two contributions (from bare $NbS_2$ and $Cr_{1/3}NbS_2/NbS_2$ stack) to the total $R(T)$.

To validate our interpretation of the data in Fig. 1, we make another $Cr_{1/3}NbS_2/NbS_2$ vdW device (Fig. 2(a)), where all electrodes are in contact with the $Cr_{1/3}NbS_2/NbS_2$ stack. As shown in Fig. 2(b), the $R(T)$ of this device does not exhibit any $R_{peak}$, independently on $H$, which supports our claim that $R_{peak}$ in Fig. 1 originates from two regions with different $T_c$.

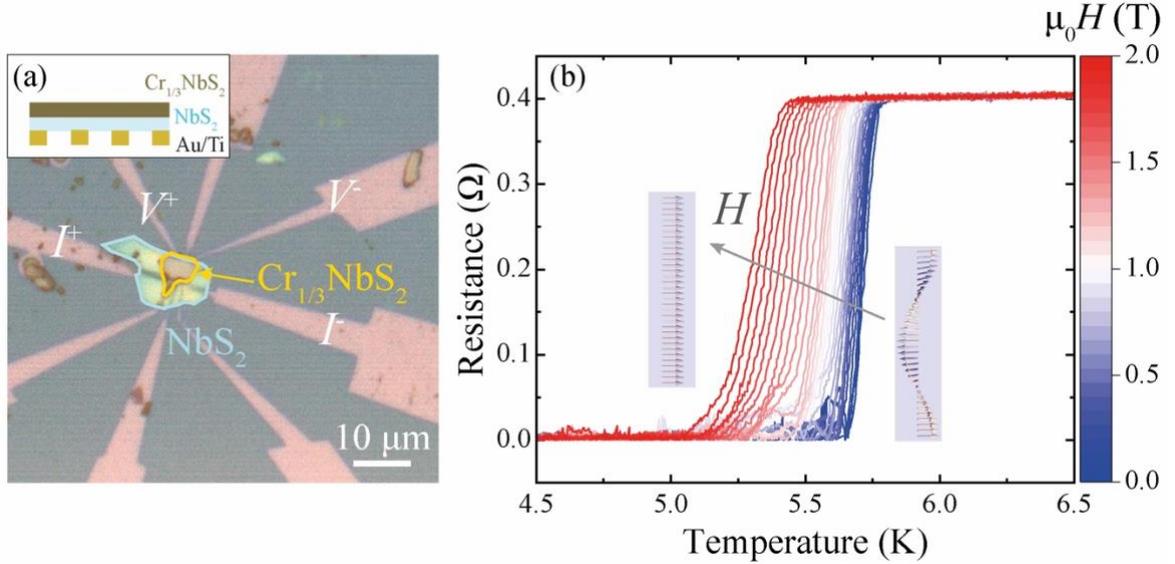

FIG. 2. (a) Optical microscope image of a $Cr_{1/3}NbS_2$ (350 nm)/$NbS_2$ (25 nm) vdW heterostructure made on Au (33 nm)/Ti (7 nm) electrodes with inset showing the materials stack and corresponding $R(T)$ in (b) as a function of an in-plane $H$ (values in the color bar) used to switch $Cr_{1/3}NbS_2$ from a helimagnetic to a ferromagnetic state.

To determine if long-ranged spin triplets affect the $T_{c,bi}$ of the device in Fig. 2(a), from each of the $R(T)$ curves in Fig. 2(b), we extract $T_{c,bi}$ defined as the $T$ where $R$ reaches the 50% of its normal-state value at $T = 6.0$ K ($T_c \sim 5.75$ K). Based on these $T_{c,bi}$ values, we study how $T_{c,bi}$ varies with $H$, $T_{c,bi}(H)$, for both increasing and decreasing $H$ (Fig. 3(a)). We always apply $H$ in the normal state of our devices at $T = 7.0$ K and, at every $H$, we record $R(T)$ whilst both cooling (from 7.0 K down to 3.0 K) and warming (from 3.0 K to 7.0 K) our samples, to extract the corresponding $T_{c,bi}$ values. Unless differently specified, the $T_{c,bi}(H)$ curves shown are based on sample-cooling $T_{c,bi}$ data.

$T_{c,bi}(H)$ for the device in Fig. 2(a) is shown in Fig. 3(a), for both $H$ upsweep (blue curve) and downsweep (red curve). As $H$ is increased, $T_{c,bi}(H)$ shows a first feature (at $\mu_0 H$

< 60 mT) followed by a decrease with different slope at 60 mT < $\mu_0 H$ < $\mu_0 H_{sat}$ and then by a positive shift at $H = H_{sat}$. During the $H$ downsweep, for $H > H_{sat}$, $T_{c,bi}(H)$ coincides with $T_{c,bi}(H)$ measured during the $H$ upsweep, but shows a difference of ~ 20 mK from the latter for $H < H_{sat}$.

The positive shift in $T_{c,bi}(H)$ at $H \sim H_{sat}$ in Fig. 3(a) is consistent with our explanation for the data in Fig. 1 and the disappearance of $R_{peak}$ at $H \sim H_{sat}$ due to a positive $T_{c,bi}$ shift. Also for this device in Fig. 3(a), if long-ranged spin triplets are generated, then $T_{c,bi}(H)$ in the Cr$_{1/3}$NbS$_2$ helimagnetic state ($H < H_{sat}$) must be lower than $T_{c,bi}(H)$ in the Cr$_{1/3}$NbS$_2$ ferromagnetic state ($H > H_{sat}$), meaning that $T_{c,bi}$ should exhibit a positive shift at $H \sim H_{sat}$ – as we observe. Consistently with the disappearance of $R_{peak}$ in Fig. 1(b), the jump in $T_{c,bi}(H)$ at $H \sim H_{sat}$ in Fig. 3(a) cannot be due to a conventional F/S proximity involving short-ranged Cooper pairs or due to stray fields, since both effects should decrease rather than increase $T_{c,bi}$ at $H \sim H_{sat}$. Vortices also cannot account for our results, as discussed in the Supplementary Material.

To rule out stray fields, like in previous studies on thin film multilayers [15,31], we fabricate a vdW device where a 10-nm-thick insulator (I) of hexagonal boron nitride (hBN) is placed between Cr$_{1/3}$NbS$_2$ and NbS$_2$ to suppress any superconducting proximity effects. The $T_c$ profile of this F/I/S vdW heterostructure (Supplementary Fig. S4), $T_{c,tri}(H)$, shows some features at $\mu_0 H$ < 100 mT like those in Fig. 3(a) for $\mu_0 H$ < 60 mT, which we associate to stray fields, but then follows a parabolic trend different from $T_{c,bi}(H)$ in Fig. 3(a). This parabolic trend resembles that measured for bare NbS$_2$ (S) (Supplementary Fig. S5). Also, $T_{c,tri}(H)$ has a negative rather than positive jump at $H \sim H_{sat}$, unlike $T_{c,bi}(H)$ in Fig. 3(a), confirming that stray fields should get stronger and reduce $T_{c,bi}$ at $H \sim H_{sat}$. Based on these results for the F/I/S device, we conclude that only the features in $T_{c,bi}(H)$ at $\mu_0 H$ < 60 mT in Fig. 3(a) are due to stray fields, whilst all the other features including the positive jump at $H \sim H_{sat}$ are due to a proximity effect involving long-ranged spin triplets.

Spectroscopic studies on Nb/Ho thin films have shown that spin-triplet generation is sensitive to the misalignment of the magnetic moments near the Nb/Ho interface, compared to the bulk Ho helix [12]. Like for Nb/Ho [12], interface magnetic moments misaligned to the Cr$_{1/3}$NbS$_2$ helix can have a non-null out-of-plane component and generate dipolar fields into NbS$_2$, which can account for the $T_{c,bi}(H)$ variation at low $H$ in our S/F vdW devices.

Previous transport experiments on epitaxial Ho/Nb thin film heterostructures [32] also show, consistently with our results, that $T_{c,bi}$ decreases as Ho is driven from its helimagnetic to ferromagnetic state by an in-plane $H$. However, when Ho becomes ferromagnetic, a negative shift in $T_{c,bi}$ is observed which, as argued by the authors, is opposite to that expected for spin-triplet generation. By contrast, in our F/S vdW devices, we observe a positive shift in $T_{c,bi}(H)$ at $H \sim H_{sat}$.

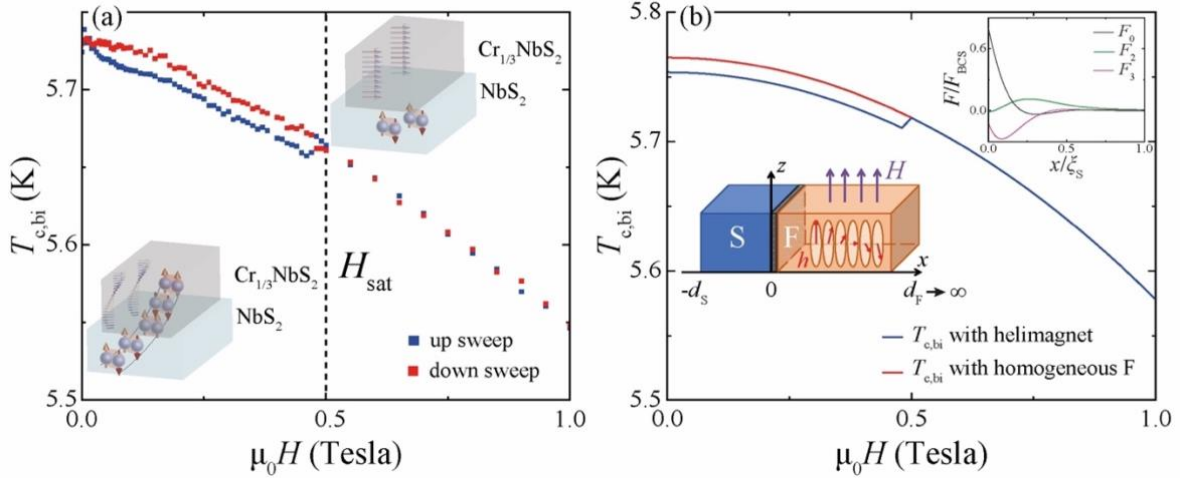

FIG. 3. (a) $T_c$ versus in-plane $H$ of a $Cr_{1/3}NbS_2/NbS_2$ device, $T_{c,bi}(H)$, determined from $R(T)$ curves in Fig. 2(b) for $H$ upsweep (blue curve) and downsweep (red curve). The lower and upper sketches show the magnetic configuration of $Cr_{1/3}NbS_2$ for $H < H_{sat}$ and $H > H_{sat}$, respectively. (b) Theoretical $T_{c,bi}(H)$ for a S/F with a helimagnet becoming ferromagnetic at $H_{sat}$ (blue curve) and for S/F with homogenous ferromagnet (red curve). The theoretical model used is shown in the lower panel, whilst the upper inset shows the decay of pairing amplitudes from the S/F interface in the F helimagnetic state: spin-singlets ($F_0$), long-ranged triplets ($F_2$) and short-ranged triplets ($F_3$).

In addition to the F/S vdW device in Fig. 3(a), we have made another F/S vdW device with thinner $NbS_2$ (~ 10 nm), to determine whether the shift in $T_{c,bi}$ increases as $d_S$ gets reduced and closer to $\xi_S$ – which would further indicate that the $T_{c,bi}(H)$ shift is due to a superconducting proximity effect. As shown in Supplementary Fig. 5(a), this device has two regions, one made of a $Cr_{1/3}NbS_2/NbS_2$ stack and the other made of bare $NbS_2$ (S) but of the same $NbS_2$ flake used for the F/S stack. Each of these regions has its own set of electrodes to measure $T_{c,bare}(H)$ independently on $T_{c,bi}(H)$. For this device, we find that $T_{c,bare}(H)$ has a parabolic trend consistent with $H$-induced orbital depairing (Supplementary Fig. 5(b)). In contrast, $T_{c,bi}(H)$ deviates from $T_{c,bare}(H)$ for $H < H_{sat}$ and shows a first drop at $\mu_0 H < 80$ mT, which we ascribe to stray fields, followed by a slower decrease that we relate to spin-triplet generation (see above). For $H > H_{sat}$, where long-ranged spin triplets cannot be generated, $T_{c,bi}(H)$ follows $T_{c,bare}(H)$, but with a constant (negative) offset due to $h_{ex}$. For 80 mT $< \mu_0 H < \mu_0 H_{sat}$, the proximity effect must therefore involve long-ranged spin triplets, since a F/S proximity with short-ranged triplets, which occurs for $H > H_{sat}$, gives a $T_{c,bi}(H)$ profile like $T_{c,bare}(H)$ (see theory model below). Also, the shift in $T_{c,bi}$ from $T_{c,bare}$ for this device is ~ 40 mK at $H < H_{sat}$, which is larger than that in Fig. 3(a), consistently with the decrease in NbS$_2$ thickness from 20 nm to 10 nm.

To support our claims that long-ranged spin-triplets are generated at the Cr$_{1/3}$NbS$_2$/NbS$_2$ vdW interface, we develop a model describing the proximity effect in our F/S system. Our theoretical description of the effect is based on the quasiclassical Green's function method in the diffusive limit (Usadel approach). A sketch of our model is reported in the inset of Fig. 3(b). We consider a S of length $d_S$ coupled to an infinitely-long helimagnet F with $d_F \gg \xi_F$. The in-plane magnetization (in the $y$-$z$ plane) of the helimagnet is modeled using the exchange field $\boldsymbol{h_{ex}} = h_{ex}(\boldsymbol{e}_y \sin qx + \boldsymbol{e}_z \cos qx)$, where $q$ relates to the pitch of the helix ($2\pi/q$) defined along the $x$-direction perpendicular to the S/F interface. The quality of the interface in our model is related to the parameter $\gamma_B$ (see Supplementary Material), which defines the interface transparency, and to $\gamma = \sigma_F/\sigma_S$ depending on the mismatch between the normal-state conductivities of F ($\sigma_F$) and S ($\sigma_S$). We assume that an in-plane $H$ is applied (along the $z$-direction), which modulates the F's magnetic textures and penetrates the S affecting $T_{c,bi}$, and we only consider $H$-induced orbital depairing. Also, we simply assume that the transition from a helimagnetic to a ferromagnetic state in F occurs abruptly at $H_{sat}$ like for the device in Fig. 3(a) and we model it as a step function.

Our hypothesis is that an inhomogeneous (helical) magnetization in a F can generate long-range spin-triplet correlations at the S/F interface thus suppressing the superconducting gap in S. Effectively, this gap suppression is seen in our theoretical curves as a reduction of the superconducting critical temperature $T_{c,bi}$. We observe that the theoretical $T_{c,bi}(H)$ in Fig. 3(b) (blue curve) calculated with our model for a helimagnet becoming ferromagnetic at $\mu_0 H_{sat} \sim 0.5$ T is in very good agreement with the experimental data in Fig. 3(a). In the same Fig. 3(b), we also report the theoretical $T_{c,bi}(H)$ profile (red curve) calculated for a S/F system with a homogeneous $h_{ex}$ confirming a parabolic trend.

The inset of Fig. 3(b) shows the anomalous Green's functions (pairing amplitudes) as a function of the position $x$ inside F. We find that the helical ordering of the F induces a non-vanishing long-ranged pairing amplitude, which suppresses the S gap and reduces $T_{c,bi}$. The curves in Fig. 3(b) are calculated using the following parameters: $d_S = 2\,\xi_S$, $\gamma = 0.055$, $\gamma_B = 0$, $q\xi_S = 3$, $h/\Delta_0 = 100$, $\xi_S = 10$ nm ($\Delta_0$ being the NbS$_2$ superconducting gap). The bulk $T_c$ and upper critical field of S are assumed to be $T_{c,S} = \Delta_0/(1.764 k_B) = 6.3$ K ($k_B$ being the Boltzmann constant) and $\mu_0 H_{c,S} = 5$ T, respectively. The only fitting parameters are $\gamma$ and $\gamma_B$, whereas all the other parameters are taken from the literature or estimated. The best fits for $T_{c,bi}(H)$ are obtained for $\gamma_B = 0$, which suggests that the F/S vdW interface must have a good transparency, to observe our experimental results. This confirms that we can get a strong proximity effect in our devices despite the weak vdW nature of their F/S interfaces.

In conclusion, the dependence of $T_{c,bi}$ in $Cr_{1/3}NbS_2$/$NbS_2$ devices on the magnetic state of $Cr_{1/3}NbS_2$ is consistent with the generation of long-ranged spin-triplet pairs across their vdW interface. A strong proximity effect involving long-ranged spin-triplet pairs can therefore not only occur across a S/F interface with strong covalent bonds, as already reported, but also across a weakly-bonded vdW interface. Our results pave the way for the combination of helimagnetic metals like $Cr_{1/3}NbS_2$ or stacks of vdW ferromagnets with other vdW superconductors to study how spin-triplet pairs can be generated and manipulated in vdW systems for superconducting spintronics.

**Acknowledgements**

We thank David Mandrus for providing the $Cr_{1/3}NbS_2$ single crystals during the first phase of the project and Uli Nowak and Moumita Kundu for discussion. A.S., H.S., A.D.B, W.B., and E.S. acknowledge funding from the Deutsche Forschungsgemeinschaft (DFG) through the SPP 2244 priority programme (grant No. 443404566). O.M. acknowledges support from the Academia Sinica – Hebrew University Research program and the Harry de Jur Chair in Applied Science. We also acknowledge support from the EU's Horizon 2020 research and innovation program under Grant Agreement No. 964398 (SUPERGATE) and from DFG via SFB 1432 (project No. 425217212).

**Conflicts of interest**

There are no conflicts to declare.